\documentclass[aps,pre,superscriptaddress,twocolumn]{revtex4}

\usepackage{graphicx,amssymb,amsfonts,amsmath,chemarr,color,commath}

\begin{document}

\title{Physical constraints on accuracy and persistence during {{breast}} {{cancer}} cell chemotaxis}

\author{Julien Varennes\footnote{These authors contributed equally to this work.}}
\affiliation{Department of Physics and Astronomy, Purdue University, West Lafayette, IN 47907, USA}

\author{Hye-ran Moon\footnotemark[1]}
\affiliation{School of Mechanical Engineering, Purdue University, West Lafayette IN 47907, USA}

\author{Soutick Saha}
\affiliation{Department of Physics and Astronomy, Purdue University, West Lafayette, IN 47907, USA}

\author{Andrew Mugler}
\email{amugler@purdue.edu}
\affiliation{School of Mechanical Engineering, Purdue University, West Lafayette IN 47907, USA}
\affiliation{Purdue Center for Cancer Research, Purdue University, West Lafayette, IN 47907, USA}

\author{Bumsoo Han}
\email{bumsoo@purdue.edu}
\affiliation{School of Mechanical Engineering, Purdue University, West Lafayette IN 47907, USA}
\affiliation{Purdue Center for Cancer Research, Purdue University, West Lafayette, IN 47907, USA}

\begin{abstract}
Directed cell motion in response to an external chemical gradient occurs in many biological phenomena such as wound healing, angiogenesis, and cancer metastasis. Chemotaxis is often characterized by the accuracy, persistence, and speed of cell motion, but whether any of these quantities is physically constrained by the others is poorly understood. Using a combination of theory, simulations, and 3D chemotaxis assays on {{single}} metastatic breast cancer cells, we investigate the links among these different aspects of chemotactic performance. In particular, we observe in both experiments and simulations that the chemotactic accuracy, but not the persistence or speed, increases with the gradient strength. We use a random walk model to explain this result and to propose that cells' chemotactic accuracy and persistence are mutually constrained. Our results suggest that key aspects of chemotactic performance are inherently limited regardless of how favorable the environmental conditions are.
\end{abstract}

\maketitle

\section*{AUTHOR SUMMARY}
One of the most ubiquitous and important cell behaviors is chemotaxis: the ability to move in the direction of a chemical gradient. Due to its importance, key aspects of chemotaxis have been quantified for a variety of cells, including the accuracy, persistence, and speed of cell motion. However, whether these aspects are mutually constrained is poorly understood. Can a cell be accurate but not persistent, or vice versa? Here we use theory, simulations, and experiments on cancer cells to uncover mutual constraints on the properties of chemotaxis. Our results suggest that accuracy and persistence are mutually constrained.

\section*{INTRODUCTION}

Chemotaxis plays a crucial role in many biological phenomena such as organism development, immune system targeting, and cancer progression \cite{iglesias2008navigating, roussos2011chemotaxis, kim2013cooperative, varennes2016sense}. Specifically, recent studies indicate that chemotaxis occurs during metastasis in many different types of cancer \cite{roussos2011chemotaxis, friedl2011cancer, witsch2010roles, woodhouse1997general, wang2004differential, shields2007autologous}. At the onset of metastasis, tumor cells invade the surrounding extracellular environment, and oftentimes chemical signals in the environment can direct the migration of invading tumor cells. Several recent experiments have quantified chemotaxis of tumor cells in the presence of different chemoattractants \cite{kim2013cooperative} and others have been devoted to the intracellular biochemical processes involved in cell motion \cite{petrie2009random}. Since the largest cause of death in cancer patients is due to the metastasis, it is important to understand and prevent the directed and chemotactic behavior of invading tumor cells.

Chemotaxis requires sensing, polarization, and motility \cite{shi2013interaction}. A cell's ability to execute these interrelated aspects of chemotaxis determines its performance. High chemotactic performance can be defined in terms of several properties. Cell motion should be accurate: cells should move in the actual gradient direction, not a different direction. Cell motion should be persistent: cells should not waste effort moving in random directions before ultimately drifting in the correct direction. Cell motion should be fast: cells should arrive at their destination in a timely manner.

\begin{figure}
\centering
\includegraphics[width=85 mm]{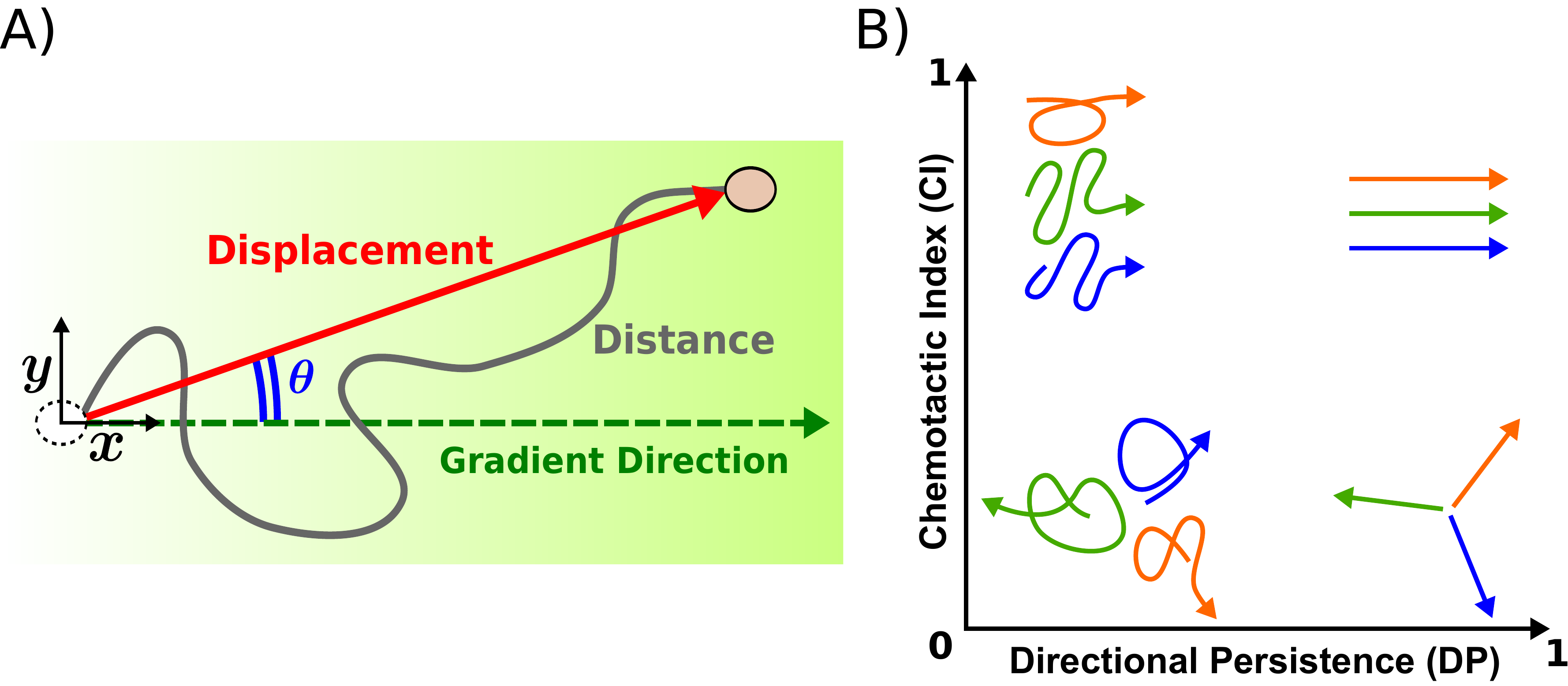}
\caption{{\bf Illustration of chemotaxis.} (A) The cell's displacement makes an angle $\theta$ with the gradient direction. The chemotactic index (CI) is defined here as the ratio of the displacement in the gradient direction to the total displacement. The {{directional persistence (DP)}} is defined here as the ratio of the total displacement to the total distance traveled. (B) {{High CI values are indicative of  cell movement in the gradient direction, whereas high {{DP}} values are indicative of straight cell movement in any direction.}}}
\label{intro}
\end{figure}

Indeed, most studies of chemotaxis use one or more of these measures to quantify chemotactic performance. Accuracy is usually quantified by the so-called chemotactic index (CI), most often defined in terms of the angle made with the gradient direction \cite{funamoto2001role, mouneimne2006spatial, van2007biased, kay2008changing} (Fig \ref{intro}A); although occasionally it is defined in terms of the ratio of distances traveled \cite{nelson1975chemotaxis} or number of motile cells \cite{iellem2001unique, mayr2002vascular, fiedler2005vegf} in the presence vs.\ absence of the gradient. {Directional persistence \cite{petrie2009random} (DP)} is usually quantified by the ratio of the magnitude of the cell's displacement (in any direction) to the total distance traveled by the cell (Fig \ref{intro}A; sometimes called the McCutcheon index \cite{mccutcheon1946chemotaxis}, length ratio \cite{gorelik2014quantitative}, or straightness index \cite{codling2008random}), although recent work has pointed out advantages of using the directional autocorrelation time \cite{gorelik2014quantitative, dang2013inhibitory}. Speed is usually quantified in terms of instantaneous speed along the trajectory or net speed over the entire assay.

However, the relationship among the accuracy, persistence, and speed in chemotaxis, and whether one quantity constrains the others, is not fully understood. Are there cells that are accurate but not very persistent, or persistent but not very accurate (Fig \ref{intro}B)? If not, is it because such motion is possible but not fit, or is it because some aspect of cell motion fundamentally prohibits this combination of chemotactic properties?

Here we focus on how a cell's intrinsic migration mechanism as well as properties of the external environment place constraints on its chemotactic performance. The physics of diffusion places inherent limits on a cell's ability to sense chemical gradients \cite{varennes2017emergent}. These limits, along with the cell's internal information processing and its motility mechanism, determine the accuracy, persistence, and speed of migration. Using a human breast cancer cell line (MDA-MB-231) embedded within a 3D collagen matrix inside a microfluidic device imposing a chemical gradient, we are able to quantify the chemotactic performance of invasive cancer cells in response to various chemical concentration profiles. Results from chemotaxis assays are then compared with simulations and theoretical predictions in order to probe the physical limits {{of cancer cells}} to chemotaxis.

\section*{RESULTS}

\subsection*{Quantifying accuracy, persistence, and speed}

We measure accuracy using the chemotactic index (CI) \cite{funamoto2001role, mouneimne2006spatial, van2007biased, kay2008changing}
\begin{equation}
\label{eq:CI1}
\text{CI} \equiv \langle \cos \theta \rangle,
\end{equation}
where $\theta$ is the angle the cell's displacement makes with the gradient direction (Fig \ref{intro}A), and the average is taken over many cell trajectories. CI is bounded between $-1$ and $1$. For chemotaxis in response to an attractant, as in this study, CI generally falls between $0$ and $1$; whereas in response to a repellent, CI usually falls between $-1$ and $0$. $\text{CI} = 1$ represents perfectly accurate chemotaxis in which cell displacement is parallel to the gradient direction (Fig \ref{intro}B, top two examples), and $\text{CI} = 0$ indicates that the cells' migration is unbiased (Fig \ref{intro}B, bottom two examples). The facts that CI is bounded and dimensionless make it easy to compare different values across different experimental conditions, and get an intuitive picture for the type of cell dynamics it represents.

We measure persistence using the {{directional persistence (DP)}}, defined as the ratio of the magnitude of the cell's displacement {{(in any direction)}} to the total distance traveled \cite{mccutcheon1946chemotaxis, gorelik2014quantitative, codling2008random} (Fig \ref{intro}A),
\begin{equation}
\label{eq:CR1}
\text{{{DP}}} \equiv \left\langle \frac{|\text{displacement}|}{\text{distance}} \right\rangle.
\end{equation}
{{Note that this ratio goes by several names \cite{mccutcheon1946chemotaxis, gorelik2014quantitative, codling2008random}, and although the name we use here contains the word `chemotactic,' the ratio is in fact independent of the gradient direction. Indeed, {{DP}} measures the tendency of a cell to move in a straight line, in any direction.}}
{{DP}} is also dimensionless and bounded between $0$ and $1$, and once again intuitive sense can be made of either limit. If $\text{{{DP}}} = 1$, then the cells are moving in perfectly straight lines {{in any arbitrary direction}} (Fig \ref{intro}B, right two examples). In contrast, {{a low {{DP}}}} is representative of a cell trajectory that starts and ends near the same location on average (Fig \ref{intro}B, left two examples), {{with {{DP}} $\to 0$ in the limit of an infinitely long non-persistent trajectory.}}

An alternative measure of persistence is the directional autocorrelation time $\tau_{\rm AC} = \int_0^\infty dt'\ \langle \cos(\theta_{t+t'}-\theta_t)\rangle$, where $t'$ is the time difference between two points in a trajectory, and the average is taken over all starting times $t$ \cite{gorelik2014quantitative, dang2013inhibitory}. The advantage of the autocorrelation time is that, unlike the {{DP}}, it is largely independent of the measurement frequency and total observation time. The disadvantage is that, unlike the {{DP}}, it is not dimensionless or bounded. Although we use the {{DP}} here, we verify in Fig S1 that the autocorrelation time varies monotonically with the {{DP}} for our experimental assay.

We measure speed using the instantaneous speed along the trajectory. That is, we take the distance traveled in the measurement interval $\Delta t$ ($15$ minutes in the experiments, see below), divide it by the interval, and average this quantity over all intervals that make up the trajectory.

\begin{figure}
\centering
\includegraphics[width=85mm]{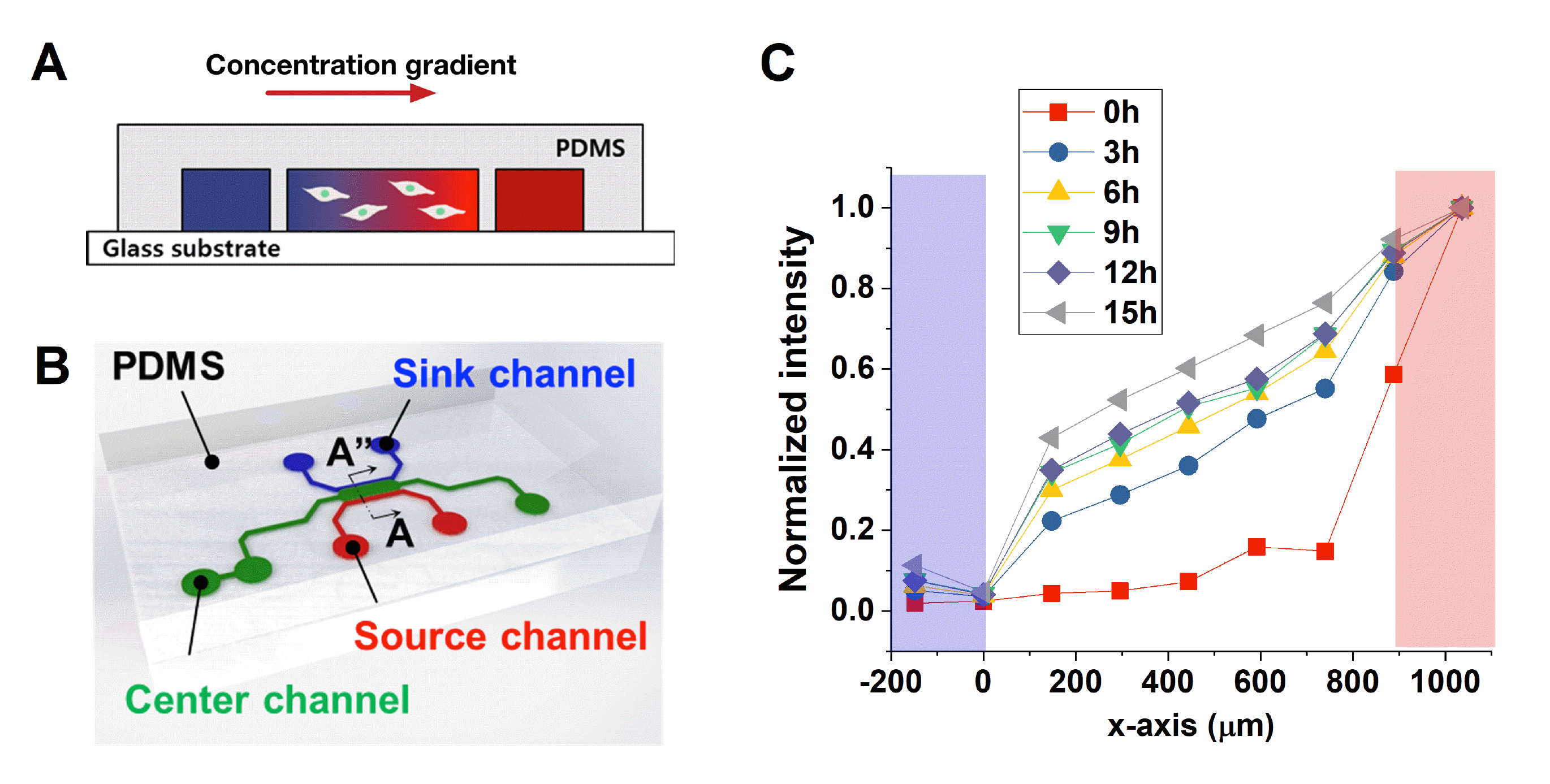}
\caption{{\bf Microfluidic device used as a chemotaxis platform.} (A) Cross-sectional view illustrating concentration gradient formed by diffusion. (B) Illustration showing structure of the microfluidic channels. Center channel (green) is filled with type I collagen mixture and MDA-MB-231 mixture, source channel is filled with culture medium containing TGF-$\beta$, and sink channel is filled with only culture medium. (C) FITC-dextran fluorescence within the center channel. Blue region indicates sink channel while red region indicates source channel.}
\label{device}
\end{figure}

\subsection*{Breast cancer cells chemotax up TGF-$\beta$ gradients}

We begin by investigating the above properties of chemotaxis in the context of metastasis, specifically the epithelial-mesenchymal transition and subsequent invasion of cancer cells. To this end, we perform experiments using a triple-negative human breast cancer cell line (MDA-MB-231). Invasion of tumor cells {\it in vivo} is aided by external cues including soluble factors that are thought to form gradients in the tumor microenvironment \cite{roussos2011chemotaxis, friedl2011cancer, witsch2010roles, woodhouse1997general, wang2004differential, shields2007autologous}. Among these soluble factors, transforming growth factor-$\beta$ (TGF-$\beta$) is a key environmental cue for the invasion process \cite{roussos2011chemotaxis, luwor2015single, giampieri2009localized, ikushima2010tgfbeta, kleuser200817}. Therefore, we use TGF-$\beta$ as the chemoattractant.

The {\it in vivo} tumor microenvironment is highly complex. As a result, {\it in vitro} platforms have been developed and widely used to investigate the cancer response to a specific cue. In this study, a microfluidic platform is used to expose the TGF-$\beta$ gradient to the cells in 3D culture condition (Fig \ref{device}A). The microfluidic device is designed with three different channels, a center, source, and sink channel (Fig \ref{device}B). The center channel is filled with a composition of MDA-MB-231 cells and type I collagen while the medium is perfused through the side source and sink channels. TGF-$\beta$ is applied only through the source channel, not the sink channel, and therefore a graded profile develops over time in the center channel by diffusion. Consequently, the MDA-MB-231 cells surrounded by type I collagen are exposed to a chemical gradient of TGF-$\beta$.

To verify that a graded TGF-$\beta$ profile is generated in the center channel, we utilize 10kDa FITC-dextran,
whose hydrodynamic radius ($2.3$ nm) is similar to that of TGF-$\beta$ (approximately $2.4$ nm \cite{venturoli2005ficoll}). The fluorescence intensity is shown in Fig \ref{device}C. The profile approaches steady state within 3 hours, is approximately linear, and remains roughly stationary for more than 12 hours. Therefore, we record the MDA-MB-231 cells using time-lapse microscopy every 15 minutes from 3 to 12 hours after imposing the TGF-$\beta$. See Materials and methods for details.

\begin{figure}
\centering
\includegraphics[width=85mm]{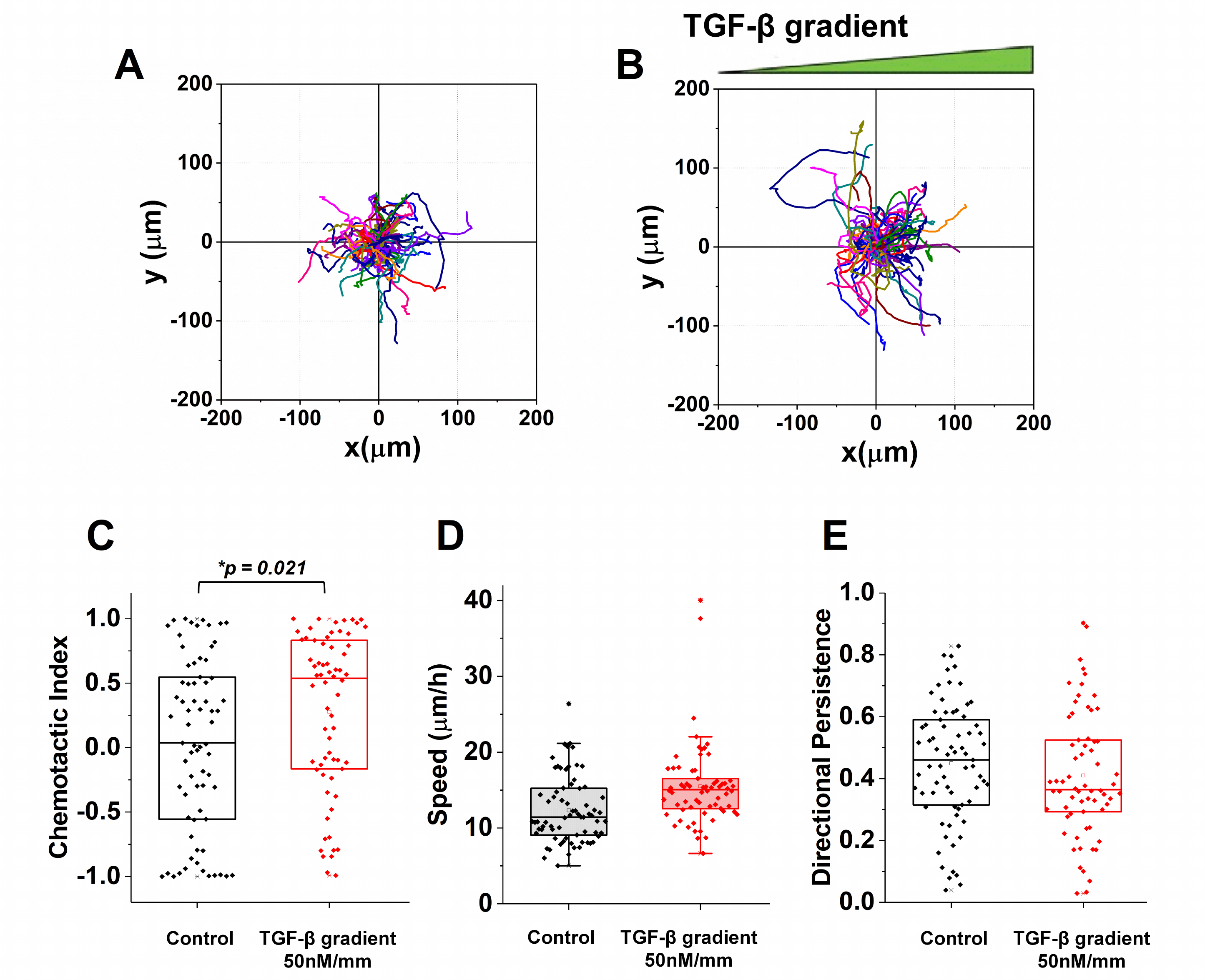}
\caption{{\bf Cell trajectories and chemotaxis metrics.} Cell trajectories of (A) control and (B) 50nM/mm TGF-$\beta$ gradient. Distribution of (C) chemotactic index, (D) speed, and (E) {{directional persistence}} of each trajectory from both the control (black) and the TGF-$\beta$ gradient (red). Boundary of box plots indicates quadrants with centerline as median. Distributions are statistically compared using Mann-Whitney test.}
\label{expt}
\end{figure}

First, we perform a control experiment with no TGF-$\beta$ to characterize the baseline of the MDA-MB-231 cell migratory behavior. Representative trajectories are shown in Fig \ref{expt}A, and we see that there is no apparent preferred direction. Indeed, as seen in Fig \ref{expt}C (black), the CI is centered around zero, indicating no directional bias. Notably, the spread of the CI values is very broad, with many data points falling near the endpoints $-1$ and $1$. This is a generic feature of the CI due to its definition as a cosine: when the distribution of angles $\theta$ is uniform, the distribution of $\cos\theta$ is skewed toward $-1$ and $1$ because of the cosine's nonlinear shape. Nonetheless, we see that the median of the CI is very near zero as expected. The speed and {{DP}} are shown in Fig \ref{expt}D and E, respectively (black). We see that the {{DP}} is significantly above zero, indicating that even in the absence of any chemoattractant, cells exhibit persistent motion. This result is consistent with previous works that showed that cells cultured in 3D tend to have directionally persistent movement unlike those in 2D \cite{petrie2009random}.

Next, we expose cells to a TGF-$\beta$ gradient of $g = 50$ nM/mm. Representative trajectories are shown in Fig \ref{expt}B, and we see a possible bias in the gradient direction. Indeed, as seen in Fig \ref{expt}C (red), the CI is centered above zero, indicating a directional bias, and the difference with the control distribution is statistically significant ($p$ value $< 0.05$). We also see in Fig \ref{expt}D (red) that the speed increases, although we will see below that the increase is relatively small and that the trend is non necessarily monotonic. Finally, we see in Fig \ref{expt}E (red) that the {{DP}} decreases, although the difference with the control is not statistically significant. These results suggest that a TGF-$\beta$ gradient causes a significant increase in directional bias (CI) but not necessarily a significant change in cell speed or persistence ({{DP}}).

\begin{figure}
\centering
\includegraphics[width=85mm]{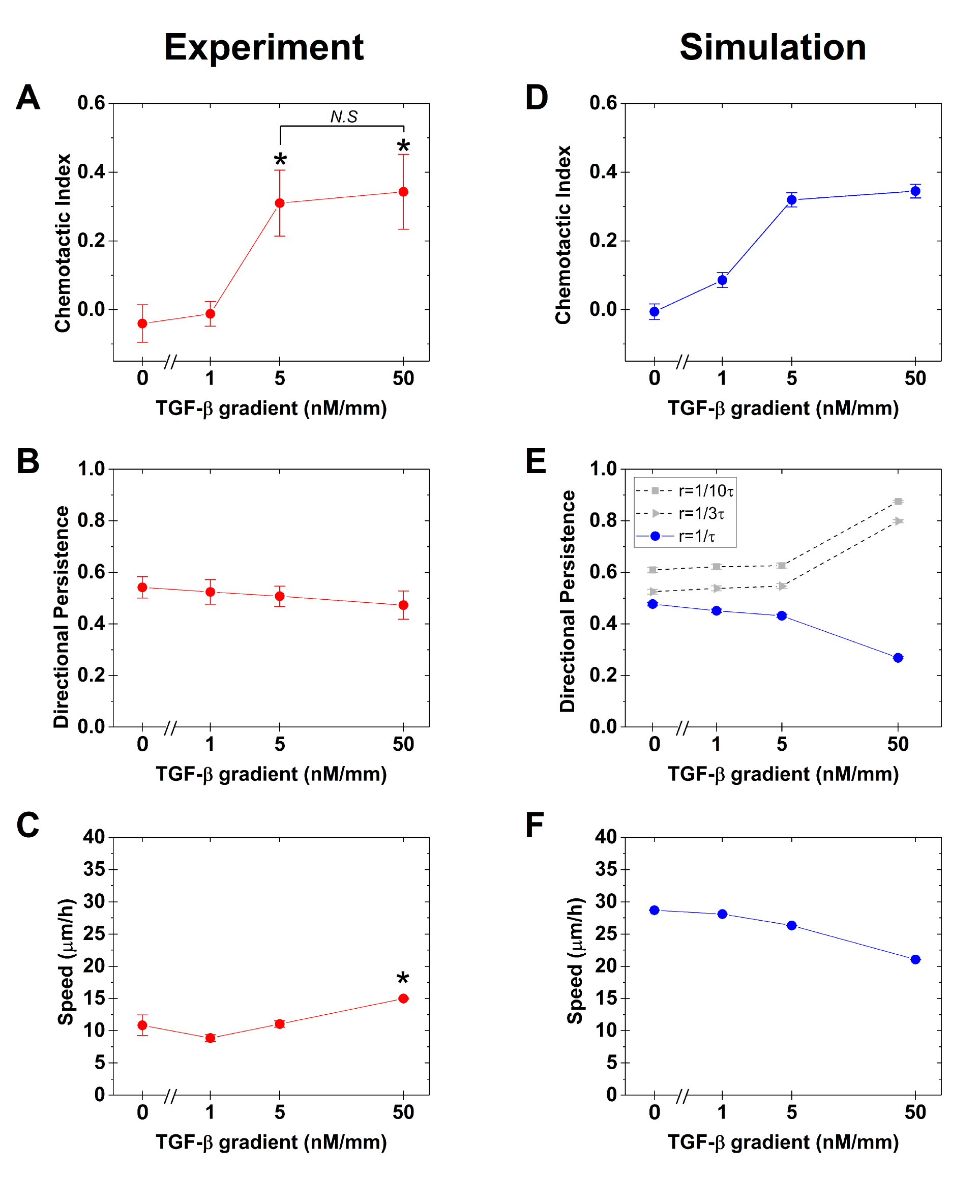}
\caption{{\bf Comparison of experiments with simulations.} Experimental (A) chemotactic index, (B) {{directional persistence}}, and (C) speed for four different TGF-$\beta$ gradients, $g = 0$, $1$, $5$, and $50$ nM/mm.{{(red)}} Data points indicate average and standard error of medians from three different experiments. A, B, and C are plotted with log-scaled TGF-$\beta$ gradient. (D-F) Same for cellular Potts model (CPM) simulations {{(blue). Error bars are standard error from 1000 trials. {{Directional persistence}} from reduced polarization memory decay rate(r) is represented in (E) (gray)}}}
\label{compare}
\end{figure}

To confirm the trends suggested above, we evaluate the response to four different TGF-$\beta$ gradient strengths, $g = 0$, $1$, $5$, and $50$ nM/mm, in three separate experiments each (Fig \ref{compare}A-C; the trajectories for all experiments and $g$ values are shown in Fig S2). We see in Fig \ref{compare}A that, consistent with Fig \ref{expt}, the CI is zero for the control and increases with gradient strength $g$. In fact, the CI appears to saturate beyond $5$ nM/mm, such that its value at $50$ nM/mm is not significantly larger than its value at $5$ nM/mm. We also see in Fig \ref{compare}B, consistent with Fig \ref{expt}, the {{DP}} slightly decreases with the gradient strength although the decrease is roughly within error bars. Finally, we see in Fig \ref{compare}C that the increase in the speed is small, achieving a statistically significant difference with the control only at the largest gradient strength, and that the trend is not monotonic.

\subsection*{Minimum detectable gradient is shallow}

A striking feature of Fig \ref{compare}A is that the cells respond to a gradient as shallow as $g = 5$ nM/mm. To put this value in perspective, we estimate both the relative concentration change and the absolute molecule number difference across the cell body \cite{varennes2016sense}. The microfluidic device is about $1$ mm in the gradient direction, and therefore a cell in the middle experiences a background concentration of about $c = 2.5$ nM. Assuming the cell is on the order of $a = 10$ $\mu$m wide, the change in concentration across its body is $ga = 0.05$ nM, for a relative change of $ga/c = 2\%$. The number of attractant molecules that would occupy half the cell body is on the order of $ca^3 = 1500$. Two percent of this is $ga^4 = 30$, meaning that cells experience about a thirty-molecule difference between their two halves. The same quantities are approximately $ga/c = 1\%$ and $6\%$, and $ga^4 = 60$ and $300$, for amoebae in cyclic
adenosine monophosphate gradients \cite{van2007biased} and epithelial cells in epidermal growth factor gradients \cite{ellison2016cell}, respectively \cite{varennes2016sense}. This suggests that the response of MDA-MB-231 cells to TGF-$\beta$ gradients is close to the physical detection limit for single cells.

\subsection*{Simulations suggest sensing and persistence are decoupled}

To understand the experimental observation that the CI increases with gradient strength, but the {{DP}} and speed do not (Fig \ref{compare}A-C), we turn to computer simulations.
{{The cells in the experiments are executing 3D migration through the collagen matrix (as opposed to crawling on top of a 2D substrate). Nevertheless, the imaging is acquired as a 2D projection of the 3D motion. We do not expect this projection to introduce much error into the analysis because the height of the microfluidic device is less than 100 $\mu$m, whereas its width in the gradient direction is about $1$ mm, and its length is several millimeters. Indeed, from the experimental trajectories (Fig \ref{expt}) we have estimated that if motility fluctuations in the height direction are equivalent to those in the length direction, then the error in the CI that we make by the fact that we only observe a 2D projection of cell motion is less than 1\%. Consequently, for simplicity we use a 2D rather than 3D simulation of chemotaxis of a cell through an extracellular medium.}}

\begin{figure}
\centering
\includegraphics[width=85mm]{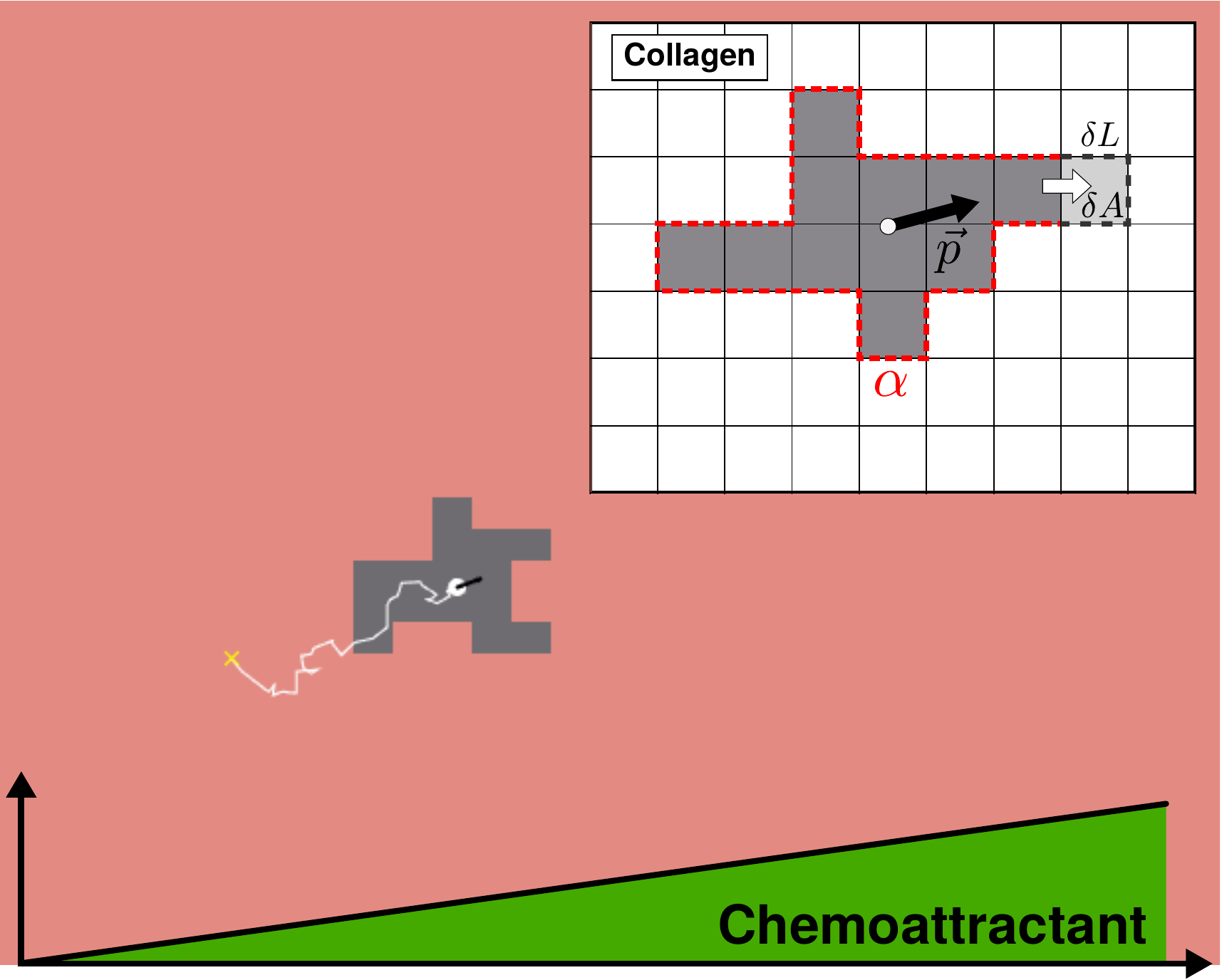}
\caption{{\bf Cellular Potts model (CPM) simulation.} Snapshot shows cell (gray) migrating towards increasing chemical concentration over time (white trajectory). Inset: Cell motility occurs through addition and removal of lattice sites. $\vec{p}$, cell polarization vector; $\alpha$, cell-collagen adhesion energy; {$\delta L$, change in perimeter; $\delta A$, change in area.}}
\label{cpm}
\end{figure}

Specifically, we use the cellular Potts model (CPM) \cite{graner1992simulation, swat2012multi}, a lattice-based simulation that has been widely used to model cell migration \cite{szabo2010collective, kabla2012collective, varennes2016collective} {{(note that whereas often the CPM is used to model collective migration, here we use it for single-cell migration)}}. In the CPM, a cell is defined as a finite set of simply connected sites on a regular square lattice (Fig \ref{cpm}). The cell adheres to the surrounding collagen with an adhesion energy $\alpha$ and has a basal area $A_0$ from which it can fluctuate at an energetic cost $\lambda$. This gives the energy function
\begin{equation}
\label{eq:CPMu1}
u = \alpha L + \lambda(A - A_0)^2,
\end{equation}
where $L$ and $A$ are the cell's perimeter and area, respectively.

Cell motion is a consequence of minimizing the energy $u$ subject to thermal noise and a bias term $w$ that incorporates the response to the gradient \cite{szabo2010collective}. Specifically, for a lattice with $S$ total sites, one update step occurs in a fixed time $\tau$ and consists of $S$ attempts to copy a random site's label (cell or non-cell) to a randomly chosen neighboring site. Each attempt is accepted with probability
\begin{equation}
\label{eq:ch2_prob}
\mathcal{P} =
	\begin{cases}
		e^{-\left( \Delta u - w \right)} &\ \Delta u - w > 0 \\
		1 &\ \Delta u - w \leq 0,
	\end{cases}
\end{equation}
where $\Delta u$ is the change in energy associated with the attempt. The bias term is defined as
\begin{equation}
w = \Delta \vec{x} \cdot \vec{p},
\end{equation}
where $\Delta \vec{x}$ is the change in the cell's center of mass caused by the attempt, and $\vec{p}$ is the cell's polarization vector (Fig \ref{cpm} inset, black arrow), described below. The dot product acts to bias cell motion because movement parallel to the polarization vector results in a more positive $w$, and thus a higher acceptance probability (Eq.\ \ref{eq:ch2_prob}).

The polarization vector is updated every time step $\tau$ according to
\begin{equation}
\label{eq:CPMp1}
\frac{\Delta \vec{p}}{\tau}= r(-\vec{p}+ \eta\Delta\hat{x}_\tau + \epsilon \vec{q}).
\end{equation}
The first term in Eq.\ \ref{eq:CPMp1} represents exponential decay of $\vec{p}$ at a rate $r$. Thus, $r^{-1}$ characterizes the polarization vector's memory timescale. The second term causes alignment of $\vec{p}$ with $\Delta\hat{x}_\tau$ according to a strength $\eta$, where $\Delta\hat{x}_\tau$ is a unit vector pointing in the direction of the displacement of the center of mass in the previous time step $\tau$. Thus, this term promotes persistence because it aligns $\vec{p}$ in the cell's previous direction of motion. The third term causes alignment of $\vec{p}$ with $\vec{q}$ according to a strength $\epsilon$, where $\vec{q}$ contains the gradient sensing information, as defined below. Thus, this term promotes bias of motion in the gradient direction.

The sensing vector $\vec{q}$ is an abstract representation of the cell's internal gradient sensing network and is defined as
\begin{equation}
\label{eq:CPMq1}
\vec{q} = \langle(n_i-\bar{n})\hat{r}_i\rangle,
\end{equation}
where the average is taken over all lattice sites $i$ that comprise the cell, {{and receptor saturation is incorporated as described below.}} The unit vector $\hat{r}_i$ points from the cell's center of mass to site $i$, the integer $n_i$ represents the number of TGF-$\beta$ molecules detected by receptors at site $i$, and $\bar{n}$ is the average of $n_i$ over all sites. The integer $n_i$ is the minimum of two quantities: (i) the number of TGF-$\beta$ receptors at site $i$, which is sampled from a Poisson distribution whose mean is the total receptor number $N$ divided by the number of sites; and (ii) the number of TGF-$\beta$ molecules in the vicinity of site $i$, which is sampled from a Poisson distribution whose mean is $(c+gx_i)\ell^3$, where $\ell$ is the lattice spacing, and $x_i$ is the position of site $i$ along the gradient direction. Taking the minimum incorporates receptor saturation, since each site cannot detect more attractant molecules than its number of receptors. The subtraction in Eq.\ \ref{eq:CPMq1} makes $\vec{q}$ a representation of adaptive gradient sensing: if receptors on one side of the cell detect molecule numbers that are higher than those on the other side, then $\vec{q}$ will point in that direction. {{Adaptive sensing has been observed in the TGF-$\beta$ pathway \cite{frick2017sensing} in the form of fold-change detection \cite{shoval2010fold} (for shallow gradients, subtraction as in Eq.\ \ref{eq:CPMq1} is similar to taking a ratio as in fold-change detection \cite{ellison2016cell}).}}

The simulation is performed at a fixed background concentration $c$ and gradient $g$ for a total time $T$. The position of the cell's center of mass is recorded at time intervals $\Delta t$, from which we compute the CI, {{DP}}, and speed.

The parameter values used in the simulation are listed in Table \ref{table1} and are set in the following way. The values $T=9$ h, $\Delta t = 15$ min, $c=2.5$ nM, and $g=5$ nM/mm are taken from the experiments. We estimate $A_0 = 400$ $\mu$m$^2$ from the experiments, and we take $\ell = 2$ $\mu$m, such that a cell typically comprises $A_0/\ell^2 = 100$ lattice sites. We find that realistic cell motion is sensitive to $\alpha$: when $\alpha$ is too small the cell is diffuse and unconnected, whereas when $\alpha$ is too large the cell does not move because the cost of perturbing the perimeter is too large. The crossover occurs around $\alpha \sim \ell^{-1}$ as expected, and therefore we set $\alpha$ on this order, to $\alpha = 2$ $\mu$m$^{-1}$. In contrast, we find that cell motion is not sensitive to $\lambda$ (apart from $\lambda = 0$ for which the cell evaporates), and therefore we set $\lambda = 0.01$ $\mu$m$^{-4}$ corresponding to typical area fluctuations of $\lambda^{-1/2}/A_0 = 2.5\%$. In order for our Poisson sampling procedure to be valid, the time step $\tau$ must be much larger than the timescale $\ell^2/D$ for an attractant molecule or receptor to diffuse with coefficient $D$ across a lattice site. Taking $D \sim 10$ $\mu$m$^2$/s, we find $\tau \gg 0.4$ s. At the other end, we must have $\tau < \Delta t = 900$ s for meaningful data collection. We find that within these bounds, results are not sensitive to $\tau$, and therefore we set $\tau$ on the larger end at $\tau = 100$ s to reduce computational run time.

\begin{table}[ht]
\begin{center}
\resizebox{\columnwidth}{!}{%
\begin{tabular}{ |l|l|l| }
\hline
Parameter & Value & Reason \\ \hline \hline
Total time $T$ & $9$ h & Experiments \\ \hline
Recording interval $\Delta t$ & $15$ min & Experiments \\ \hline
Background concentration $c$ & $2.5$ nM & Experiments \\ \hline
Concentration gradient $g$ & $5$ nM/mm & Experiments \\ \hline
Relaxed cell area $A_0$ & $400 \ \mu\text{m}^2$ & Experiments \\ \hline
Lattice spacing $\ell$ & $2$ $\mu$m & $\sim$$100$ sites per cell \\ \hline
Cell-environment contact energy $\alpha$ & $2$ $\mu$m$^{-1}$ & $\alpha \sim \ell^{-1}$ \\ \hline
Area deviation energy $\lambda$ & 0.01 $\mu$m$^{-1}$ & $\lambda^{-1/2} \ll A_0$ \\ \hline
Simulation time $\tau$ & 100 s & $\ell^2/D \ll \tau < \Delta t$ \\ \hline
Total receptor number $N$ & 10,000 & CI saturation \\ \hline
Bias strength $\epsilon$ & 56 & Calibrated via CI \\ \hline
Persistence strength $\eta$ & 107 & Calibrated via {{DP}} \\ \hline
Polarization memory decay rate $r$ & 0.01 s$^{-1}$ & $r \sim \tau^{-1}$ \\ \hline
\end{tabular}
}
\caption{Table of parameters and values used in cellular Potts model (CPM) simulations. See text for more detailed reasoning behind values.}
\label{table1}
\end{center}
\end{table}

The parameters $N$, $\eta$, and $\epsilon$ are calibrated from the experimental data in Fig \ref{compare}A-C. Specifically, $N$ sets the gradient value above which the CI saturates (see Fig \ref{compare}A) because if the gradient is large but $N$ is small, the cell quickly migrates into a region in which there are more attractant molecules than receptors at all lattice sites, and gradient detection is not possible. We find that $N =$ 10,000, which is a reasonable value for the number of TGF-$\beta$ receptors per cell \cite{wakefield1987distribution, mitchell1992characterization}, places the saturation level at roughly $g = 50$ nM/mm as in the experiments (Fig \ref{compare}D). We set $\epsilon = 56$ $\mu$m$^{-1}$ and $\eta = 107$ $\mu$m$^{-1}$ to calibrate their cognate observables, CI and {{DP}}, respectively, to the corresponding experimental values at $g=5$ nM/mm (Fig \ref{compare}D and E).

The final parameter is the memory timescale of the polarization vector, $r^{-1}$. As seen in Fig \ref{compare}E (gray), we find that the behavior of the {{DP}} depends sensitively on this timescale. When $r^{-1}$ is large, the {{DP}} increases with gradient strength. In contrast, when $r^{-1}$ is small (indeed, equal to the smallest timescale in the system, $\tau$), the {{DP}} does not increase with gradient strength, and in fact slightly decreases (Fig \ref{compare}E, blue). Because the latter behavior is consistent with the experiments (Fig \ref{compare}B), we set $r^{-1} = \tau$. We conclude that the memory timescale of MDA-MB-231 cells is very short when responding to TGF-$\beta$ gradients.

We validate the simulation in two ways, using the speed. First, we find that the magnitude of the speed in the simulations is on the same order as the speed in the experiments (Fig \ref{compare}C and F), i.e., tens of microns per hour. Second, we find that the speed shows little dependence on the gradient strength in both the simulations and the experiments: it slightly increases in Fig \ref{compare}C and slightly decreases in Fig \ref{compare}F. Considering that the speed is not calibrated directly in our simulations, these consistencies validate the CPM as a reasonable description of the cell migration in the experiments.

Our finding that the cell's memory timescale $r^{-1}$ takes its minimum value allows for the following interpretation: the parameter $r$ couples the persistence term and the sensory term in the CPM (Eq.\ \ref{eq:CPMp1}). Thus, when the memory timescale $r^{-1}$ is long, biased motion must be also persistent and vice versa. In contrast, when the memory timescale $r^{-1}$ is short, it is possible for bias to increase without increasing persistence. Therefore, the simulations suggest that the reason that CI but not {{DP}} increases with gradient strength in the experiments, is that the drivers of sensory bias and migratory persistence in the cell's internal network are decoupled from one another.

\subsection*{Theoretical model reveals performance constraints}
Our finding that bias and persistence are decoupled in the simulations allows us to appeal to a much more simplified theoretical model in order to understand and predict global constraints on chemotaxis performance. Specifically, we consider the biased persistence random walk (BPRW) model \cite{alt1980biased, othmer1988models}, in which bias and persistence enter as explicitly independent terms controlled by separate parameters. The BPRW has been shown to be sufficient to capture random and directional, but not periodic, behaviors of 3D cell migration \cite{fraley2012dimensional}. Because we do not observe periodic back-and-forth motion of cells in our experiments, we propose that the BPRW is sufficient to investigate chemotactic constraints here.

As in the simulations, we consider the BPRW model in 2D. In the BPRW model, a cell is idealized as a single point. Its trajectory consists of $M$ steps whose lengths are drawn from an exponential distribution. We take $M = T/\Delta t = 36$ as in the experiments. The probability of a step making an angle $\theta$ with respect to the gradient direction is
\begin{equation}
\label{eq:BPRW3}
P(\theta|\theta') = \underbrace{b\cos\theta}_{\rm bias}
	+ \underbrace{\frac{e^{p\cos(\theta-\theta')}}{2\pi I_0(p)}}_{\rm persistence},
\end{equation}
where $\theta'$ is the angle corresponding to the previous step. The first term incorporates the bias, with strength $b$. It is maximal when the step points in the gradient direction ($\theta = 0$) and therefore promotes bias in that direction. It integrates to zero over its range ($-\pi < \theta < \pi$) because the bias term only reshapes the distribution without adding or subtracting net probability. The second term incorporates the persistence, with strength $p$. It is a von Mises distribution (similar to a Gaussian distribution, but normalized over the finite range $-\pi < \theta < \pi$) whose sharpness grows with $p$. It is maximal at the previous angle $\theta'$ and therefore promotes persistence. The normalization factor $I_0$ is the zeroth-order modified Bessel function of the first kind.

\begin{figure}
\centering
\includegraphics[width=85mm]{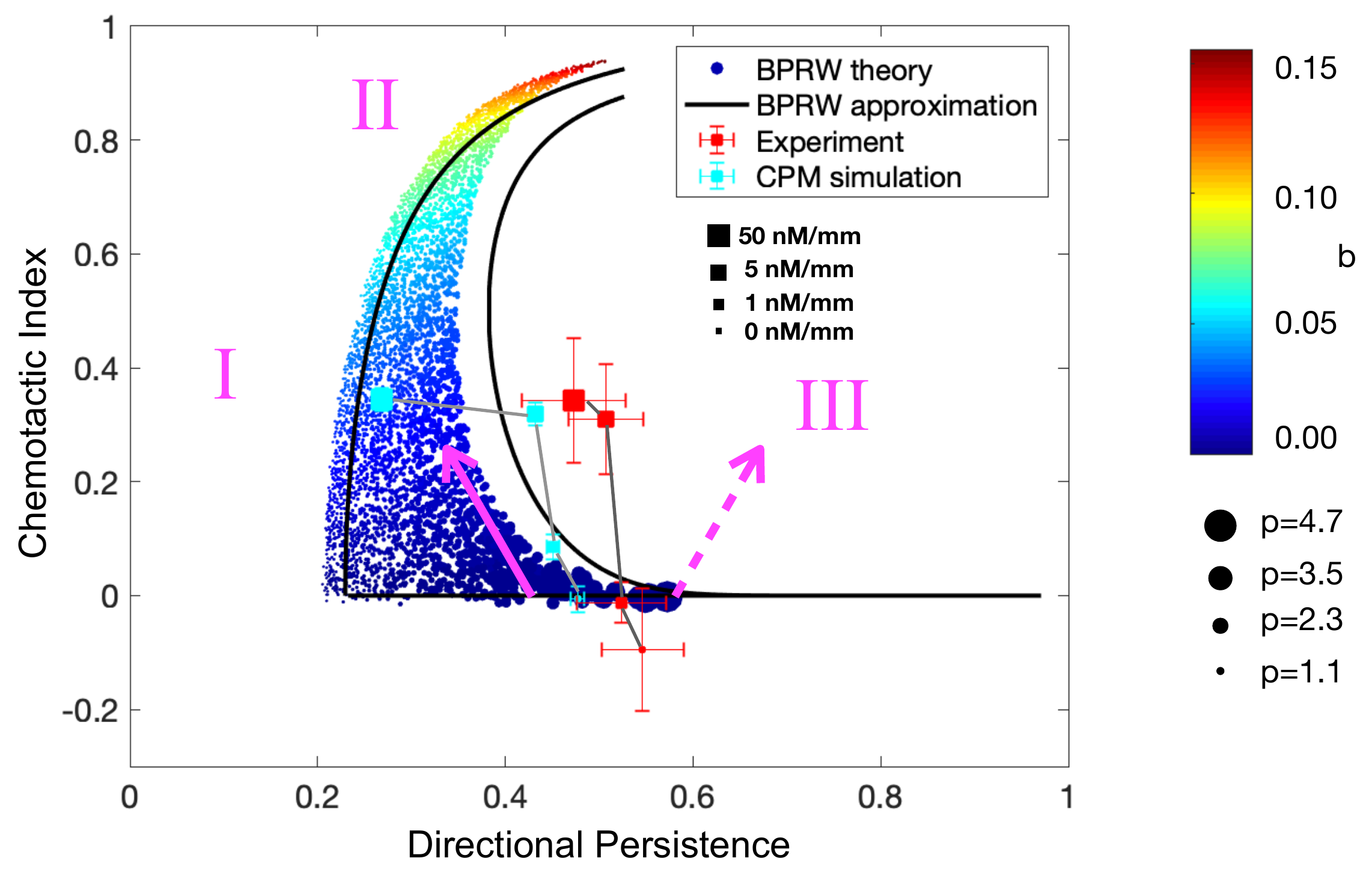}
\caption{{\bf Comparison of theory with experiments and simulations.} Colored circles show CI and {{DP}} for all values of bias parameter (color) and persistence parameter (size) for biased persistent random walk (BPRW) theory. Black lines show analytic approximations of the bounding curves. Red and cyan squares show experimental and simulation data, respectively, from Fig \ref{compare}. {{Magenta numerals and arrows show ``forbidden'' regions and qualitative trends, respectively, discussed in text.}}}
\label{bprw}
\end{figure}

The requirement that $P(\theta|\theta')$ be non-negative over the entire range of $\theta$ mutually constrains $b$ and $p$. However, apart from this constraint, $b$ and $p$ can take any positive value. We sample many pairs of $b$ and $p$, reject those that violate the constraint, and compute the CI and {{DP}} from a trajectory generated by each remaining pair. The results are shown in Fig \ref{bprw} (colored circles). We see in Fig \ref{bprw} that the BPRW model exists in a highly restricted `crescent' shape within CI--{{DP}} space. As expected, the CI increases with the bias parameter $b$ (color of circles, from blue to red). The top corner corresponds to maximal bias and no persistence; indeed, when $p=0$ the persistence term in Eq.\ \ref{eq:BPRW3} reduces to $(2\pi)^{-1}$, and non-negativity requires $b < (2\pi)^{-1} \approx 0.16$, which is consistent with the upper limit of the color bar. Also as expected, the {{DP}} increases with the persistence parameter $p$ (size of circles, from small to large), although only in the lower portion where the CI is low.

The crescent shape of the allowed CI and {{DP}} values in Fig \ref{bprw} can be understood quantitatively because several moments of the BPRW are known analytically \cite{othmer1988models}. Specifically, the mean squared displacement and the mean displacement in the gradient direction are, in units of the mean step length,
\begin{align}
\label{eq:r2}
\langle r^2\rangle &= \frac{1}{(1-\psi)^2}
	\bigg[ z^2\tilde{M}^2 + 2\left(1-2z^2-z^2e^{-\tilde{M}}\right)\tilde{M} \nonumber \\
	& \qquad\qquad\quad\,\,\,+ 2\left(2z^2-1\right)\left(1-e^{-\tilde{M}}\right) \nonumber \\
	& \qquad\qquad\quad\,\,\,+ 2z^2\left(1-e^{-\tilde{M}}\right)^2 \bigg], \\
\label{eq:x}
\langle x\rangle &= \frac{z}{1-\psi}\left(\tilde{M} - 1 + e^{-\tilde{M}}\right),
\end{align}
respectively, where $\tilde{M} = M(1-\psi)$ and $z = \chi/(1-\psi)$, with $\chi = \int_{-\pi}^\pi d\theta\ b\cos^2\theta = \pi b$ and $\psi = \int_{-\pi}^\pi d\phi\ [2\pi I_0(p)]^{-1} e^{p\cos\phi} \cos\phi = I_1(p)/I_0(p)$. We approximate the CI and {{DP}} in terms of these moments,
\begin{align}
\label{eq:CIapprox}
{\rm CI} &= \left\langle\frac{x}{r}\right\rangle \approx \frac{\langle x\rangle}{\langle r\rangle}
	\approx \frac{\langle x\rangle}{\sqrt{\langle r^2\rangle}}, \\
\label{eq:CRapprox}
{\rm {{DP}}} &= \frac{\langle r\rangle}{M} \approx \frac{\sqrt{\langle r^2\rangle}}{M},
\end{align}
and evaluate these expressions in specific limits to approximate the edges of the shape. In the limit $b=0$, Eq.\ \ref{eq:CIapprox} reduces to CI $=0$ (bottom black line in Fig \ref{bprw}). In the limit $p=0$, Eqs.\ \ref{eq:CIapprox} and \ref{eq:CRapprox} are functions of only $b$ and $M$, and $b$ can be eliminated to yield ${\rm {{DP}}} = [1+M(1-{\rm CI}^2)/2]^{-1/2}$ (left black line in Fig \ref{bprw}), where we have used the approximation $M\gg1$ (see Materials and methods).
Note here that when CI $=0$ we have {{DP}} $\approx (M/2)^{-1/2}$ for large $M$, which makes sense because for a simple random walk ($p=b=0$) the displacement goes like $M^{1/2}$ while the distance goes like $M$, such that {{DP}} $\sim M^{-1/2}$. Finally, the right edge corresponds to the maximal value of $p$ for a given $b$,
for which we compute the approximation curve parametrically (right black line in Fig \ref{bprw}; see Materials and methods). We see in Fig \ref{bprw} that these approximate expressions slightly underestimate the CI and overestimate the {{DP}}, but otherwise capture the crescent shape well. The under- and overestimation are due to the approximation $\langle r\rangle \approx \sqrt{\langle r^2\rangle}$ in Eqs.\ \ref{eq:CIapprox} and \ref{eq:CRapprox}: because $\sigma_r^2 = \langle r^2\rangle - \langle r\rangle^2 \ge 0$ for any statistical quantity, we have $\sqrt{\langle r^2\rangle} \ge \langle r\rangle$, making Eq.\ \ref{eq:CIapprox} an underestimate and Eq.\ \ref{eq:CRapprox} an overestimate.

{{The crescent shape can also be understood intuitively. First, we see that the {{DP}} cannot be smaller than a minimum value (region I in Fig \ref{bprw}). This is because the trajectory length $M$ is finite, and as discussed above, the {{DP}} only vanishes for infinitely long trajectories. If $M$ were to increase, the crescent would extend further toward {{DP}} $= 0$. Second, we see that the top of the crescent bends away from the CI $\to 1$, {{DP}} $\to 0$ corner (region II in Fig \ref{bprw}). In other words, it is not possible to have high bias without any persistence. This is because if the bias is strong, then cells will track the gradient very well. Consequently, they will move in nearly straight lines in the gradient direction, and straight movement corresponds to high persistence. This is a bias-induced persistence, distinct from the bias-independent persistence in the lower-right corner of the crescent. Finally, we see that the bending shape of the crescent implies that no solutions exist at large {{DP}} and intermediate CI (region III in Fig \ref{bprw}). In other words, it is not possible to have high persistence with partial bias. This is because, as mentioned above, persistence is induced either (i) directly, as a result of a large persistence parameter $p$ which is independent of the bias, in which case the CI is low; or (ii) indirectly, as a result of a large bias parameter $b$, in which case the CI is high. Neither of these mechanisms permits intermediate bias, and therefore high persistence can be accompanied only by low or high directionality. Together, these features of the crescent shape imply that specific modes of chemotaxis are prohibited under our simple model, as indicated by the regions I, II, and III.}}

{{Finally, the crescent shape provides a qualitative rationale for the data from the simulations and experiments,}} which are overlaid in the cyan and red squares in Fig \ref{bprw}, respectively. Specifically, the shape of the crescent is such that if a cell has a low CI and intermediate {{DP}} (bottom right corner of the crescent) and its CI increases, its {{DP}} must decrease {{(solid magenta arrow in Fig \ref{bprw}). In contrast,}} a simultaneous increase in CI and {{DP}} from this starting position is not possible according to the model {{(dashed magenta arrow in Fig \ref{bprw}). We see that the data are qualitatively consistent with this predicted trend,}} as an increase in the CI corresponds to a decrease in the {{DP}} in both the experiments and the simulations (Fig \ref{bprw}, squares). {{There is quantitative disagreement, in the sense that the data do not quite overlap with the crescent, but this is a reflection of the extreme simplicity of the BPRW model. Nonetheless, the qualitative features}} of the BPRW model are sufficient to explain the way in which accuracy and persistence are mutually constrained during the chemotaxis response of these cells.

\section*{DISCUSSION}

By integrating experiments with theory and simulations, we have investigated mutual constraints on the accuracy (CI), persistence ({{DP}}), and speed of {{cancer}} cell motion in response to a chemical attractant. We have found that while the CI of breast cancer cells increases with the strength of a TGF-$\beta$ gradient, the speed does not show a strong trend, and the {{DP}} slightly decreases. The simulations suggest that the decrease in {{DP}} is due to a decoupling between sensing and persistence in the migration dynamics. The theory confirms that the decrease in {{DP}} is due to a mutual constraint on accuracy and persistence for this type of decoupled dynamics, and more generally, it suggests that entire regions of the accuracy--persistence space are prohibited.

The present results provide some insights into TGF-$\beta$ induced migration mechanisms. Multiple signaling pathways induced by TGF-$\beta$ affect the dynamics of actin polymerization regulating cell migratory behaviors\cite{ikushima2010tgfbeta,derynck2003smad-dependent,moustakas2008dynamic,olson2010linking}. Among these, phosphatidylinositol 3-kinase (PI3K) and the small GTPase-Rac1 signaling have been reported to promote actin organization of breast cancer cells in response to TGF-$\beta$ \cite{olson2010linking,dumont2003autocrine}. PI3K and the Rho-family GTPase networks (including Rac1, RhoA and Cdc42) have been widely studied in chemotaxis, which regulates cell polarity and directional sensing \cite{krause2014steering,swaney2010eukaryotic,edlund2002transforming,fukata2003roles}. The PI3K activity, thus, can possibly explain the present chemotactic responses of the breast cancer cells to TGF-$\beta$ gradient. Recent studies have shown that PI3K is relevant to the accuracy of the cell movement in shallow chemoattractants, whereas it does not induce the orientation of cell movement in steep gradients; rather, PI3K contributes the motility enhancement \cite{andrew2007chemotaxis,bosgraaf2008pi3}. These results can be correlated with the cell motility trend in the present experimental results. In addition, the PI3K signaling pathway has been reported not to mediate the persistence of cell protrusions which could be directly related to the {{DP}} \cite{krause2014steering,swaney2010eukaryotic}. The directional persistence could be more relevant to the polarity stability which is hardly controlled by chemotaxis \cite{krause2014steering} as presented in the present results. In TGF-$\beta$ molecular cascades, activation of SMAD proteins could also affect the actin dynamics. Since SMAD-cascades include negative feedback inhibiting Rho activity\cite{derynck2003smad-dependent,moustakas2008dynamic}, it may affect the cell responses highly promoted in CI but not in speed.  However, the underlying molecular mechanisms need further research.

{{Our finding that sensing and persistence are largely decoupled in the migration dynamics is related to the view that directional sensing and polarity are separate but connected modules in chemotaxis \cite{shi2013interaction}. Indeed, CI, DP, and speed in our study play the roles of the directional sensing, polarity, and motility modules, respectively, that have been shown to reproduce many of the observed behaviors of chemotaxing cells. Moreover, several of the the molecular signaling pathways discussed above, including those involving PI3K and Rho family GTPases, have been proposed as the potential networks corresponding to these modules \cite{shi2013interaction}.}}

Several predictions arise from our work that would be interesting to test in future experiments. First, our simulation scheme assumes that the saturation of the CI with gradient strength (Fig \ref{compare}A) is due to limited receptor numbers. However, alternative explanations exist that are independent of the receptors, such as the fact that it is more difficult to detect a concentration difference on top of a large concentration background than on top of a small concentration background due to intrinsic fluctuations in molecule number \cite{mugler2016limits, ellison2016cell}. An interesting consequence of our mechanism of receptor saturation is that, at very large gradients (beyond those of Fig \ref{compare}A), the CI would actually decrease because all receptors would be bound. It would be interesting to test this prediction in future experiments.

Second, our work suggests that not all quadrants of the accuracy--persistence plane are possible for cells to achieve (Fig \ref{bprw}). It would be interesting to measure the CI and {{DP}} of other cell types, in other chemical or mechanical environments, to see if the crescent shape seen in Fig \ref{bprw} is a universal restriction, or if not, what new features of chemotaxis are therefore not captured by the modeling. In this respect, the work here can be seen as a null model, deviations from which would indicate new and unique types of cell motion.


\section*{MATERIALS AND METHODS}

\subsection*{Cell culture and reagents}
Human breast adenocarcinoma cells (MDA-MB-231) were cultured in Dulbecco’s Modified Eagle Medium/Ham’s F-12 (Advanced DMEM/F-12, Lifetechnologies, CA, USA) supplemented by 5$\%$ v/v fetal bovin serum (FBS), 2 mM L-glutamine (L-glu), and 100 \(\mu\)g ml\textsuperscript{-1}  penicillin/streptomycin(P/S) for less than 15 passages. MDA-MB-231 cells were regularly harvested by 0.05\(\%\) trypsin and 0.53mM EDTA (Lifetechnologies, CA, USA) when grown up to around 80\(\%\) confluency in 75 cm$^2$ T-flasks at 37 $^{\circ}$C with 5\(\%\) CO$_2$ incubation. Harvested cells were used for experiments or sub-cultured.

Cell-matrix composition was prepared in the microfluidic device. For the composition, MDA-MB-231 cells were mixed with 2 mg/ml of type I collagen (Corning Inc., NY, USA) mixture prepared with 10X PBS, NaOH, HEPE solution, FBS, Glu, P/S, and cell-culture level distilled water after centrifuged with 1000 rpm for 3 minutes. The cell mixture was filled in center-channel of the microfluidic devices and incubated in at 37 $^{\circ}$C with 5$\%$ CO$_2$. The cells in the collagen matrix were initially cultured in basic medium (DMEM/F12 supplemented by 5$\%$ v/v FBS, 2 mM L-glu, and 100 \(\mu\)g ml$^{-1}$ p/s) for 24 hours. Then the cells were exposed by reduced serum medium for another 24 hours, which was advanced DMEM/F12 containing 1$\%$ v/v FBS, 2 mM L-glu, and 100 \(\mu\)g ml$^{-1}$ p/s \cite{rhee2010promigratory}. After 24 hour-serum starvations, cells were exposed by a gradient of transforming growth factor beta-1 (TGF-\(\beta\)1, Invitrogen, CA, USA).

\subsection*{Microfluidic device for chemical gradient}
The microfluidic device was designed to generate a linear gradient of soluble factors (Fig \ref{device}). The device is composed of three channels which are 100 $\mu$m in thickness as described previously \cite{shin2013development}. A center channel that is 1 mm wide aims to culture tumor cells with ECM components. The center channel is connected to two side channels. The 300 $\mu$m-wide side channels are connected to large reservoirs at the end ports including culture medium. Since the side channels are in contact with the top and bottom sides of the center channel, the growth factor gradient can be generated by diffusing the soluble factor from one of the side channels, a source channel, to the other, a sink channel. Assuming there is neither pressure difference nor flow between the side channels, the concentration of a given factor can be described by the chemical species conservation equation as follows:
 
\begin{equation}
    \frac{\partial c_i}{\partial t}= {D_i} \cdot \nabla{c_i} 
\end{equation}

Once the concentration profile in the center channel reaches steady state, the linear profile persists for a while and can therefore be approximated by assuming the boundary conditions of concentration at the side channels are constants. To verify the diffusion behavior, the gradient formation was examined by using 10k Da FITC-fluorescence conjugated dextran (FITC-dextran). FITC-dextran solution was applied in the source channel while the sink channel was filled with normal culture medium. The FITC-dextran concentration profile was evaluated by the FITC fluorescent intensity in the center channel. To disregard the effect of photo-bleaching on the results, the intensity was normalized by the intensity of the source channel. The normalized intensity was reasonably considered since the fluorescence intensity of the source channel consistently remained as maximum due to the large reservoirs. The FITC dextran intensity profile (Fig \ref{device}C) showed that the linear profile was developed within 3 hours after applying the source and continued for more than 9 hours.


\subsection*{Characterization of cell migration with time-lapse microscopy}
	
Cell behaviors were captured every 15 minutes for 9 hours using an inverted microscope (Olympus IX71, Japan) equipped with a stage top incubator as described previously \cite{ozcelikkale2017effects, ozcelikkale2017differential,shin2016characterization}, so that the microfluidic platform could be maintained at 37 $^{\circ}$C in a 5\% CO$_2$ environment during imaging. The time-lapse imaging was started 3 hours after applying TGF-$\beta$1 solution in the source channel to have sufficient adjusting time. To analyze each cell behavior, a cell area in the bright field images were defined by a contrast difference between the cells and a background, and the images were converted to monochrome images by using ImageJ. Cell trajectories were demonstrated by tracking centroids of the cell area. In tracking the cell movements, cells undergoing division were excluded to avoid extra influences to affect cell polarity \cite{harley2008microarchitecture}. Moreover, stationary cells due to the presence of the matrix were excluded \cite{harley2008microarchitecture, giampieri2009localized, clark2015modes, haessler2012migration}. The stationary cells were defined as the cells that moved less than their diameter. A migration trajectory was defined by connecting the centroids of a cell from each time point.

\subsection*{Statistical analysis of experiments}

In examining the chemotactic characteristics of each group, more than 40 cell trajectories were evaluated per a group. A data point in Fig \ref{expt}C-E indicates each metric of a cell trajectory showing distribution characteristics with a box plot. The box plot includes boundaries as quadrants and a center as a median. The distribution of each metric was statistically analyzed by using Mann-Whitney U-test. This non-parametric method was used since the distribution was not consistently normal (the CI is a function of cosine). The significant change on the population lies on the biased distribution of each cell parameter when the $p$ value $<0.05$. Furthermore, the experiments were repeated at least 3 times and reported with means of medians $\pm$ standard estimated error (S.E.M.) in Fig \ref{compare}A-C. To evaluate physical limits on each metric, the data points were compared each other using a student t-test. The statistical significance between comparisons were examined when the $p$ value $<0.05$.

\subsection*{Mathematical approximations}
In the limit $p=0$, Eqs.\ \ref{eq:r2} and \ref{eq:x} become
\begin{eqnarray}
\langle r^2\rangle &=& z^2M^2 + 2(1-2z^2)M + 2(3z^2-1), \\
\langle x\rangle^2 &=& z^2(M-1)^2,
\end{eqnarray}
where $z = \pi b$, and we have neglected the exponential terms in the limit $M\gg1$. Defining the small parameter $\epsilon = 1/M$, these expressions become
\begin{eqnarray}
\langle r^2\rangle &=& z^2M^2(1+c\epsilon), \\
\langle x\rangle^2 &=& z^2M^2(1-2\epsilon)
\end{eqnarray}
to first order in $\epsilon$, where $c \equiv 2(z^{-2}-2)$. Inserting these expressions into Eqs.\ \ref{eq:CIapprox} and \ref{eq:CRapprox}, we obtain
\begin{eqnarray}
\label{eq:CIe}
{\rm CI}^2 &=& 1 - (c+2)\epsilon, \\
\label{eq:CRe}
{\rm {{DP}}}^2 &=& z^2(1+c\epsilon)
\end{eqnarray}
to first order in $\epsilon$. Because $z$ and $c$ are both functions only of $b$, we eliminate $b$ from Eqs.\ \ref{eq:CIe} and \ref{eq:CRe} to obtain 
\begin{equation}
{\rm CI}^2 = 1-2\epsilon\frac{1-{\rm {{DP}}}^2}{{\rm {{DP}}}^2}
\end{equation}
to first order in $\epsilon$. This expression is equivalent to that given below Eq.\ \ref{eq:CRapprox} and provides the left black line in Fig \ref{bprw}.

The right black line in Fig \ref{bprw} corresponds to the maximal value of $p$ for a given $b$ that keeps Eq.\ \ref{eq:BPRW3} non-negative. Non-negativity requires that the sum of the minimal values of each term in Eq.\ \ref{eq:BPRW3} is zero: $-b + e^{-p}/[2\pi I_0(p)] = 0$. With this expression for $b$ in terms of $p$, Eqs.\ \ref{eq:CIapprox} and \ref{eq:CRapprox} become functions of only $p$ and $M$. Therefore, by varying $p$, we compute the right black line parametrically.

\section*{Acknowledgments}
This work was supported by the Ralph W. and Grace M. Showalter Research Trust, the Purdue Research Foundation, and the Purdue University Center for Cancer Research Challenge Award.


\begin{thebibliography}{10}

\bibitem{iglesias2008navigating}
Iglesias PA, Devreotes PN.
\newblock Navigating through models of chemotaxis.
\newblock Current Opinion in Cell Biology. 2008;20(1):35--40.

\bibitem{roussos2011chemotaxis}
Roussos ET, Condeelis JS, Patsialou A.
\newblock Chemotaxis in cancer.
\newblock Nature Reviews Cancer. 2011;11(8):573--587.

\bibitem{kim2013cooperative}
Kim BJ, Hannanta-Anan P, Chau M, Kim YS, Swartz MA, Wu M.
\newblock Cooperative roles of SDF-1$\alpha$ and EGF gradients on tumor cell
  migration revealed by a robust 3D microfluidic model.
\newblock PloS ONE. 2013;8(7):e68422.

\bibitem{varennes2016sense}
Varennes J, Mugler A.
\newblock Sense and sensitivity: physical limits to multicellular sensing,
  migration, and drug response.
\newblock Molecular Pharmaceutics. 2016;13(7):2224--2232.

\bibitem{friedl2011cancer}
Friedl P, Alexander S.
\newblock Cancer invasion and the microenvironment: plasticity and reciprocity.
\newblock Cell. 2011;147(5):992--1009.

\bibitem{witsch2010roles}
Witsch E, Sela M, Yarden Y.
\newblock Roles for growth factors in cancer progression.
\newblock Physiology. 2010;25(2):85--101.

\bibitem{woodhouse1997general}
Woodhouse EC, Chuaqui RF, Liotta LA.
\newblock General mechanisms of metastasis.
\newblock Cancer. 1997;80(S8):1529--1537.

\bibitem{wang2004differential}
Wang SJ, Saadi W, Lin F, Nguyen CMC, Jeon NL.
\newblock Differential effects of EGF gradient profiles on MDA-MB-231 breast
  cancer cell chemotaxis.
\newblock Experimental Cell Research. 2004;300(1):180--189.

\bibitem{shields2007autologous}
Shields JD, Fleury ME, Yong C, Tomei AA, Randolph GJ, Swartz MA.
\newblock Autologous chemotaxis as a mechanism of tumor cell homing to
  lymphatics via interstitial flow and autocrine CCR7 signaling.
\newblock Cancer Cell. 2007;11(6):526--538.

\bibitem{petrie2009random}
Petrie RJ, Doyle AD, Yamada KM.
\newblock Random versus directionally persistent cell migration.
\newblock Nature Reviews Molecular Cell Biology. 2009;10(8):538--549.

\bibitem{shi2013interaction}
Shi C, Huang CH, Devreotes PN, Iglesias PA.
\newblock Interaction of motility, directional sensing, and polarity modules
  recreates the behaviors of chemotaxing cells.
\newblock PLoS computational biology. 2013;9(7):e1003122.

\bibitem{funamoto2001role}
Funamoto S, Milan K, Meili R, Firtel RA.
\newblock Role of phosphatidylinositol 3 kinase and a downstream pleckstrin
  homology domain--containing protein in controlling chemotaxis in
  Dictyostelium.
\newblock The Journal of Cell Biology. 2001;153(4):795--810.

\bibitem{mouneimne2006spatial}
Mouneimne G, DesMarais V, Sidani M, Scemes E, Wang W, Song X, et~al.
\newblock Spatial and temporal control of cofilin activity is required for
  directional sensing during chemotaxis.
\newblock Current Biology. 2006;16(22):2193--2205.

\bibitem{van2007biased}
Van~Haastert PJ, Postma M.
\newblock Biased random walk by stochastic fluctuations of
  chemoattractant-receptor interactions at the lower limit of detection.
\newblock Biophysical Journal. 2007;93(5):1787--1796.

\bibitem{kay2008changing}
Kay RR, Langridge P, Traynor D, Hoeller O.
\newblock Changing directions in the study of chemotaxis.
\newblock Nature Reviews Molecular Cell Biology. 2008;9(6):455--463.

\bibitem{nelson1975chemotaxis}
Nelson RD, Quie PG, Simmons RL.
\newblock Chemotaxis under agarose: a new and simple method for measuring
  chemotaxis and spontaneous migration of human polymorphonuclear leukocytes
  and monocytes.
\newblock The Journal of Immunology. 1975;115(6):1650--1656.

\bibitem{iellem2001unique}
Iellem A, Mariani M, Lang R, Recalde H, Panina-Bordignon P, Sinigaglia F,
  et~al.
\newblock Unique chemotactic response profile and specific expression of
  chemokine receptors CCR4 and CCR8 by CD4+ CD25+ regulatory T cells.
\newblock The Journal of Experimental Medicine. 2001;194(6):847--854.

\bibitem{mayr2002vascular}
Mayr-Wohlfart U, Waltenberger J, Hausser H, Kessler S, G{\"u}nther KP, Dehio C,
  et~al.
\newblock Vascular endothelial growth factor stimulates chemotactic migration
  of primary human osteoblasts.
\newblock Bone. 2002;30(3):472--477.

\bibitem{fiedler2005vegf}
Fiedler J, Leucht F, Waltenberger J, Dehio C, Brenner RE.
\newblock VEGF-A and PlGF-1 stimulate chemotactic migration of human
  mesenchymal progenitor cells.
\newblock Biochemical and Biophysical Research Communications.
  2005;334(2):561--568.

\bibitem{mccutcheon1946chemotaxis}
McCutcheon M.
\newblock Chemotaxis in leukocytes.
\newblock Physiological Reviews. 1946;26(3):319--336.

\bibitem{gorelik2014quantitative}
Gorelik R, Gautreau A.
\newblock Quantitative and unbiased analysis of directional persistence in cell
  migration.
\newblock Nature Protocols. 2014;9(8):1931--1943.

\bibitem{codling2008random}
Codling EA, Plank MJ, Benhamou S.
\newblock Random walk models in biology.
\newblock Journal of the Royal Society Interface. 2008;5(25):813--834.

\bibitem{dang2013inhibitory}
Dang I, Gorelik R, Sousa-Blin C, Derivery E, Gu{\'e}rin C, Linkner J, et~al.
\newblock Inhibitory signalling to the Arp2/3 complex steers cell migration.
\newblock Nature. 2013;503(7475):281--284.

\bibitem{varennes2017emergent}
Varennes J, Fancher S, Han B, Mugler A.
\newblock Emergent versus individual-based multicellular chemotaxis.
\newblock Physical Review Letters. 2017;119(18):188101.

\bibitem{luwor2015single}
Luwor RB, Hakmana D, Iaria J, Nheu TV, Simpson RJ, Zhu HJ.
\newblock {Single live cell TGF-$\beta$ signalling imaging: breast cancer cell
  motility and migration is driven by sub-populations of cells with dynamic
  TGF-$\beta$-Smad3 activity}.
\newblock Molecular Cancer. 2015;14(1):50.

\bibitem{giampieri2009localized}
Giampieri S, Manning C, Hooper S, Jones L, Hill CS, Sahai E.
\newblock Localized and reversible TGF$\beta$ signalling switches breast cancer
  cells from cohesive to single cell motility.
\newblock Nature Cell Biology. 2009;11(11):1287.

\bibitem{ikushima2010tgfbeta}
Ikushima H, Miyazono K.
\newblock {TGF$\beta$ signalling: a complex web in cancer progression}.
\newblock Nature Reviews Cancer. 2010;10(6):415.

\bibitem{kleuser200817}
Kleuser B, Malek D, Gust R, Pertz HH, Potteck H.
\newblock {17-$\beta$-Estradiol inhibits Transforming Growth Factor-$\beta$
  signalling and function in breast cancer cells via activation of
  Extracellular Signal-Regulated Kinase through the G protein coupled receptor
  30}.
\newblock Molecular Pharmacology. 2008;74(6):1533--1543.

\bibitem{venturoli2005ficoll}
Venturoli D, Rippe B.
\newblock Ficoll and dextran vs. globular proteins as probes for testing
  glomerular permselectivity: effects of molecular size, shape, charge, and
  deformability.
\newblock American Journal of Physiology-Renal Physiology.
  2005;288(4):F605--F613.

\bibitem{ellison2016cell}
Ellison D, Mugler A, Brennan MD, Lee SH, Huebner RJ, Shamir ER, et~al.
\newblock Cell--cell communication enhances the capacity of cell ensembles to
  sense shallow gradients during morphogenesis.
\newblock Proceedings of the National Academy of Sciences.
  2016;113(6):E679--E688.

\bibitem{graner1992simulation}
Graner F, Glazier JA.
\newblock Simulation of biological cell sorting using a two-dimensional
  extended Potts model.
\newblock Physical Review Letters. 1992;69(13):2013.

\bibitem{swat2012multi}
Swat MH, Thomas GL, Belmonte JM, Shirinifard A, Hmeljak D, Glazier JA.
\newblock Multi-scale modeling of tissues using CompuCell3D.
\newblock Methods in Cell Biology. 2012;110:325.

\bibitem{szabo2010collective}
Szab{\'o} A, {\"U}nnep R, M{\'e}hes E, Twal W, Argraves W, Cao Y, et~al.
\newblock Collective cell motion in endothelial monolayers.
\newblock Physical Biology. 2010;7(4):046007.

\bibitem{kabla2012collective}
Kabla AJ.
\newblock Collective cell migration: leadership, invasion and segregation.
\newblock Journal of The Royal Society Interface. 2012; p. rsif20120448.

\bibitem{varennes2016collective}
Varennes J, Han B, Mugler A.
\newblock Collective chemotaxis through noisy multicellular gradient sensing.
\newblock Biophysical Journal. 2016;111(3):640--649.

\bibitem{frick2017sensing}
Frick CL, Yarka C, Nunns H, Goentoro L.
\newblock Sensing relative signal in the Tgf-$\beta$/Smad pathway.
\newblock Proceedings of the National Academy of Sciences. 2017; p. 201611428.

\bibitem{shoval2010fold}
Shoval O, Goentoro L, Hart Y, Mayo A, Sontag E, Alon U.
\newblock Fold-change detection and scalar symmetry of sensory input fields.
\newblock Proceedings of the National Academy of Sciences. 2010; p. 201002352.

\bibitem{wakefield1987distribution}
Wakefield LM, Smith DM, Masui T, Harris CC, Sporn MB.
\newblock Distribution and modulation of the cellular receptor for transforming
  growth factor-beta.
\newblock The Journal of Cell Biology. 1987;105(2):965--975.

\bibitem{mitchell1992characterization}
Mitchell E, Lee K, O'Connor-McCourt M.
\newblock Characterization of transforming growth factor-beta (TGF-beta)
  receptors on BeWo choriocarcinoma cells including the identification of a
  novel 38-kDa TGF-beta binding glycoprotein.
\newblock Molecular Biology of the Cell. 1992;3(11):1295--1307.

\bibitem{alt1980biased}
Alt W.
\newblock Biased random walk models for chemotaxis and related diffusion
  approximations.
\newblock Journal of Mathematical Biology. 1980;9(2):147--177.

\bibitem{othmer1988models}
Othmer HG, Dunbar SR, Alt W.
\newblock Models of dispersal in biological systems.
\newblock Journal of Mathematical Biology. 1988;26(3):263--298.

\bibitem{fraley2012dimensional}
Fraley SI, Feng Y, Giri A, Longmore GD, Wirtz D.
\newblock Dimensional and temporal controls of three-dimensional cell migration
  by zyxin and binding partners.
\newblock Nature Communications. 2012;3:719.

\bibitem{derynck2003smad-dependent}
Derynck R, Zhang YE.
\newblock Smad-dependent and Smad-independent pathways in TGF$\beta$ family
  signalling.
\newblock Nature. 2003;425:577.

\bibitem{moustakas2008dynamic}
Moustakas A, Heldin CH.
\newblock Dynamic control of TGF-$\beta$ signaling and its links to the
  cytoskeleton.
\newblock FEBS letters. 2008;582(14):2051--2065.

\bibitem{olson2010linking}
Olson EN, Nordheim A.
\newblock Linking actin dynamics and gene transcription to drive cellular
  motile functions.
\newblock Nature Reviews Molecular Cell Biology. 2010;11(5):353--365.

\bibitem{dumont2003autocrine}
Dumont N, Bakin AV, Arteaga CL.
\newblock Autocrine transforming growth factor-$\beta$ signaling mediates
  Smad-independent motility in human cancer cells.
\newblock Journal of Biological Chemistry. 2003;278(5):3275--3285.

\bibitem{krause2014steering}
Krause M, Gautreau A.
\newblock Steering cell migration: lamellipodium dynamics and the regulation of
  directional persistence.
\newblock Nature reviews Molecular cell biology. 2014;15(9):577.

\bibitem{swaney2010eukaryotic}
Swaney KF, Huang CH, Devreotes PN.
\newblock Eukaryotic chemotaxis: a network of signaling pathways controls
  motility, directional sensing, and polarity.
\newblock Annual review of biophysics. 2010;39:265--289.

\bibitem{edlund2002transforming}
Edlund S, Landstrom M, Heldin CH, Aspenstrom P.
\newblock Transforming Growth Factor-$\beta$- induced Mobilization of Actin
  Cytoskeleton Requires Signaling by Small GTPases Cdc42 and RhoA.
\newblock Molecular Biology of the Cell. 2002;13(3):902--914.

\bibitem{fukata2003roles}
Fukata M, Nakagawa M, Kaibuchi K.
\newblock Roles of Rho-family GTPases in cell polarisation and directional
  migration.
\newblock Current Opinion in Cell Biology. 2003;15(5):590--597.

\bibitem{andrew2007chemotaxis}
Andrew N, Insall RH.
\newblock Chemotaxis in shallow gradients is mediated independently of PtdIns
  3-kinase by biased choices between random protrusions.
\newblock Nature Cell Biology. 2007;9(2):193--200.

\bibitem{bosgraaf2008pi3}
Bosgraaf L, Keizer-Gunnink I, Van~Haastert PJ.
\newblock PI3-kinase signaling contributes to orientation in shallow gradients
  and enhances speed in steep chemoattractant gradients.
\newblock J Cell Sci. 2008;121(21):3589--3597.

\bibitem{mugler2016limits}
Mugler A, Levchenko A, Nemenman I.
\newblock Limits to the precision of gradient sensing with spatial
  communication and temporal integration.
\newblock Proceedings of the National Academy of Sciences.
  2016;113(6):E689--E695.

\bibitem{rhee2010promigratory}
Rhee S, Ho CH, Grinnell F.
\newblock Promigratory and procontractile growth factor environments
  differentially regulate cell morphogenesis.
\newblock Experimental Cell Research. 2010;316(2):232--244.

\bibitem{shin2013development}
Shin CS, Kwak B, Han B, Park K.
\newblock Development of an in vitro 3D tumor model to study therapeutic
  efficiency of an anticancer drug.
\newblock Molecular Pharmaceutics. 2013;10(6):2167--2175.

\bibitem{ozcelikkale2017effects}
Ozcelikkale A, Dutton JC, Grinnell F, Han B.
\newblock Effects of dynamic matrix remodelling on en masse migration of
  fibroblasts on collagen matrices.
\newblock Journal of The Royal Society Interface. 2017;14(135):20170287.

\bibitem{ozcelikkale2017differential}
Ozcelikkale A, Shin K, Noe-Kim V, Elzey BD, Dong Z, Zhang JT, et~al.
\newblock Differential response to doxorubicin in breast cancer subtypes
  simulated by a microfluidic tumor model.
\newblock Journal of Controlled Release. 2017;266:129--139.

\bibitem{shin2016characterization}
Shin K, Klosterhoff BS, Han B.
\newblock Characterization of Cell-Type-Specific Drug Transport and Resistance
  of Breast Cancers Using Tumor-Microenvironment-on-Chip.
\newblock Molecular Pharmaceutics. 2016;13(7):2214--2223.

\bibitem{harley2008microarchitecture}
Harley BAC, Kim HD, Zaman MH, Yannas IV, Lauffenburger DA, Gibson LJ.
\newblock Microarchitecture of three-dimensional scaffolds influences cell
  migration behavior via junction interactions.
\newblock Biophysical Journal. 2008;95(8):4013--4024.

\bibitem{clark2015modes}
Clark AG, Vignjevic DM.
\newblock Modes of cancer cell invasion and the role of the microenvironment.
\newblock Current Opinion in Cell Biology. 2015;36:13--22.

\bibitem{haessler2012migration}
Haessler U, Teo JC, Foretay D, Renaud P, Swartz MA.
\newblock Migration dynamics of breast cancer cells in a tunable 3D
  interstitial flow chamber.
\newblock Integrative Biology. 2012;4(4):401--409.



\end{thebibliography}

\onecolumngrid
\appendix
\newpage
\section*{Supporting information}


\begin{figure}[!ht]
\centering
\includegraphics[width=110mm]{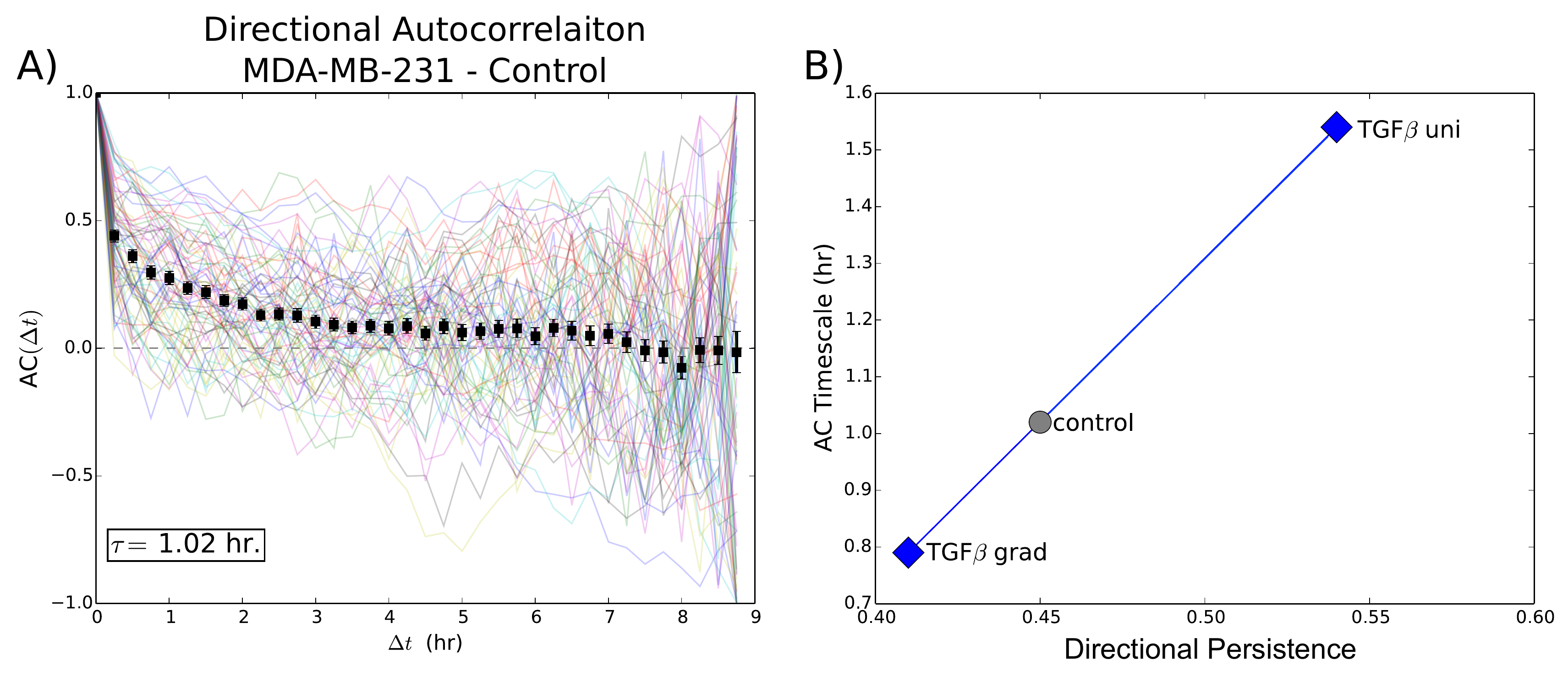}
\caption{{\bf Comparison of {{directional persistence}} ({{DP}}) and directional autocorrelation time ($\tau_{\rm AC}$).} (A) Autocorrelation function for all trajectories in control experiment (no TGF-$\beta$); $\tau_{\rm AC}$ is the integral under the curve. Plot of $\tau_{\rm AC}$ vs.\ {{DP}} for control (gray), and 50 nM/mm TGF-$\beta$ gradient condition (left blue triangle), as well as several other experimental conditions. Note that the relationship between $\tau_{\rm AC}$ and {{DP}} is monotonic.}
\label{S1_Fig}
\end{figure}

\begin{figure}[!ht]
\centering
\includegraphics[width=110mm]{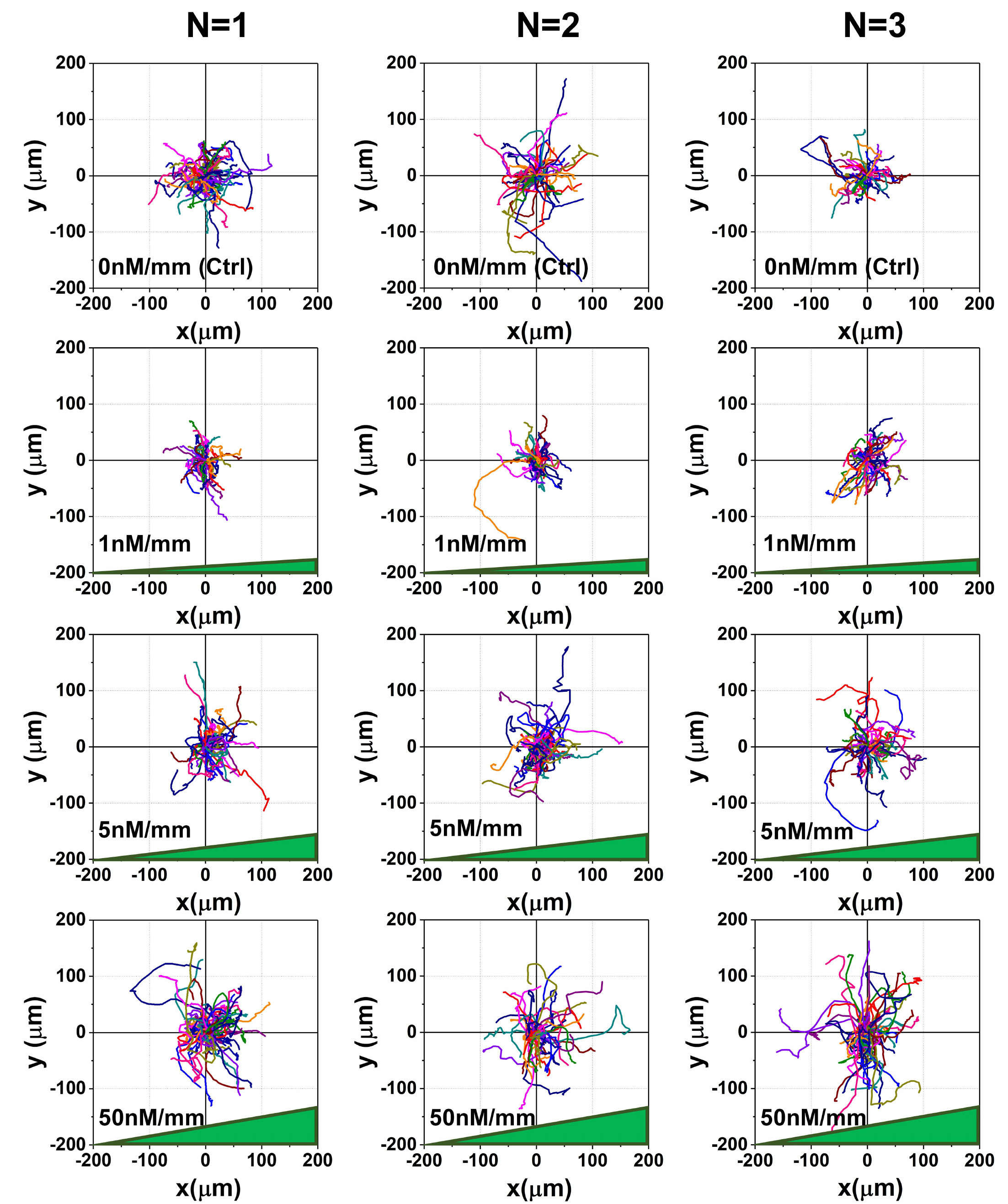}
\caption{\bf Cell trajectories for all values of TGF-$\beta$ gradient strength, and all three experimental replicates.}
\label{S2_Fig}
\end{figure}

%
%
%


\end{document}